\documentclass[aps,prl,twocolumn,showpacs,superscriptaddress,groupedaddress]{revtex4-2}  
\pdfoutput=1
\usepackage{graphicx}  
\usepackage{ragged2e}
\usepackage{dcolumn}   
\usepackage{bm}        
\usepackage{amssymb}   
\usepackage[hidelinks]{hyperref}
\usepackage{fancyhdr}
\usepackage{xcolor}
\usepackage{float}
\usepackage{lipsum}
\hypersetup{
    colorlinks,
    linkcolor={blue!80!black},
    citecolor={blue!80!black},
    urlcolor={blue!80!black}
}
\begin{document}

\title{Probing the Deuteron at Very Large Internal Momenta}

\author{C.~Yero$^{1,15}$}       
\homepage[ORCiD: ]{https://orcid.org/0000-0003-2822-7373}
\email[email: ]{cyero@jlab.org}
\author{D.~Abrams$^{2}$}           
\author{Z.~Ahmed$^{3}$}            
\author{A.~Ahmidouch$^{4}$}        
\author{B.~Aljawrneh$^{4}$}        
\author{S.~Alsalmi$^{5}$}          
\author{R.~Ambrose$^{3}$}          
\author{W.~Armstrong$^{6}$}        
\author{A.~Asaturyan$^{7}$}        
\author{K.~Assumin-Gyimah$^{8}$}   
\author{C.~Ayerbe Gayoso$^{9}$}    
\author{A.~Bandari$^{9}$}          
\author{J.~Bane$^{10}$}            
\author{S.~Basnet$^{3}$}           
\author{V.V.~Berdnikov$^{11}$}     
\author{J.~Bericic$^{15}$}         
\author{H.~Bhatt$^{8}$}            
\author{D.~Bhetuwal$^{8}$}         
\author{D.~Biswas$^{12}$}          
\author{W.U.~Boeglin$^{1}$}        
\author{P.~Bosted$^{9}$}           
\author{E.~Brash$^{13}$}           
\author{M.H.S.~Bukhari$^{14}$}     
\author{H.~Chen$^{2}$}             
\author{J.P.~Chen$^{15}$}          
\author{M.~Chen$^{2}$}             
\author{M.E.~Christy$^{12}$}       
\author{S.~Covrig$^{15}$}          
\author{K.~Craycraft$^{10}$}       
\author{S.~Danagoulian$^{4}$}      
\author{D.~Day$^{2}$}             
\author{M.~Diefenthaler$^{15}$}    
\author{M.~Dlamini$^{17}$}         
\author{J.~Dunne$^{8}$}
\author{B.~Duran$^{16}$}          
\author{D.~Dutta$^{8}$}          
\author{R.~Ent$^{15}$}            
\author{R.~Evans$^{3}$}           
\author{H.~Fenker$^{15}$}         
\author{N.~Fomin$^{10}$}          
\author{E.~Fuchey$^{18}$}         
\author{D.~Gaskell$^{15}$}        
\author{T.N.~Gautam$^{12}$}       
\author{F.A.~Gonzalez$^{19}$}     
\author{J.O.~Hansen$^{15}$}       
\author{F.~Hauenstein$^{20}$}     
\author{A.V.~Hernandez$^{11}$}    
\author{T.~Horn$^{11}$}           
\author{G.M.~Huber$^{3}$}         
\homepage[ORCiD: ]{http://orcid.org/0000-0002-5658-1065}
\author{M.K.~Jones$^{15}$}         
\author{S.~Joosten$^{6}$}          
\author{M.L.~Kabir$^{8}$}            
\author{A.~Karki$^{8}$}           
\author{C.E.~Keppel$^{15}$}          
\author{A.~Khanal$^{1}$}       
\author{P.~King$^{17}$}             
\author{E.~Kinney$^{21}$}           
\author{N.~Lashley-Colthirst$^{12}$}     
\author{S.~Li$^{22}$}               
\author{W.B.~Li$^{9}$}              
\author{A.H.~Liyanage$^{12}$}       
\author{D.J.~Mack$^{15}$}           
\author{S.P.~Malace$^{15}$}         
\author{J.~Matter$^{2}$}           
\author{D.~Meekins$^{15}$}          
\author{R.~Michaels$^{15}$}         
\author{A.~Mkrtchyan$^{7}$}        
\author{H.~Mkrtchyan$^{7}$}        
\author{S.J.~Nazeer$^{12}$}         
\author{S.~Nanda$^{8}$}           
\author{G.~Niculescu$^{23}$}        
\author{M.~Niculescu$^{23}$}        
\author{D.~Nguyen$^{2}$}           
\author{N.~Nuruzzaman$^{24}$}       
\author{B.~Pandey$^{12}$}           
\author{S.~Park$^{19}$}             
\author{C.F.~Perdrisat$^{9}$}       
\author{E.~Pooser$^{15}$}           
\author{M.~Rehfuss$^{16}$}          
\author{J.~Reinhold$^{1}$}         
\author{B.~Sawatzky$^{15}$}         
\author{G.R.~Smith$^{15}$}          
\author{A.~Sun$^{25}$}              
\author{H.~Szumila-Vance$^{15}$}    
\author{V.~Tadevosyan$^{7}$}        
\author{S.A.~Wood$^{15}$}           
\author{J.~Zhang$^{19}$}       

\collaboration{for the Hall C Collaboration}
\affiliation{$^{1}$Florida International University, University Park, Florida 33199, USA}
\affiliation{$^{2}$University of Virginia, Charlottesville, Virginia 22903, USA}
\affiliation{$^{3}$University of Regina, Regina, Saskatchewan S4S 0A2, Canada}
\affiliation{$^{4}$North Carolina Agricultural and Technical State University, Greensboro, North Carolina 27411, USA}
\affiliation{$^{5}$Kent State University, Kent, Ohio 44240, USA}
\affiliation{$^{6}$Argonne National Laboratory, Lemont, Illinois 60439, USA}  
\affiliation{$^{7}$A.I. Alikhanyan National Science Laboratory (Yerevan Physics Institute), 2 Alikhanian Brothers Street, 0036, Yerevan, Armenia}
\affiliation{$^{8}$Mississippi State University, Mississippi State, Mississippi 39762, USA}
\affiliation{$^{9}$The College of William \& Mary, Williamsburg, Virginia 23185, USA}
\affiliation{$^{10}$University of Tennessee, Knoxville, Tennessee 37996, USA}
\affiliation{$^{11}$Catholic University of America, Washington, DC 20064, USA}
\affiliation{$^{12}$Hampton University, Hampton, Virginia 23669, USA}
\affiliation{$^{13}$Christopher Newport University, Newport News, Virginia 23606, USA}
\affiliation{$^{14}$Jazan University, Jazan 45142, Saudi Arabia}
\affiliation{$^{15}$Thomas Jefferson National Accelerator Facility, Newport News, Virginia 23606, USA}
\affiliation{$^{16}$Temple University, Philadelphia, Pennsylvania 19122, USA}
\affiliation{$^{17}$Ohio University, Athens, Ohio 45701, USA}
\affiliation{$^{18}$University of Connecticut, Storrs, Connecticut 06269, USA}
\affiliation{$^{19}$Stony Brook University, Stony Brook, New York 11794, USA}
\affiliation{$^{20}$Old Dominion University, Norfolk, Virginia 23529, USA}
\affiliation{$^{21}$University of Colorado Boulder, Boulder, Colorado 80309, USA}
\affiliation{$^{22}$University of New Hampshire, Durham, New Hampshire 03824, USA}
\affiliation{$^{23}$James Madison University, Harrisonburg, Virginia 22807, USA}
\affiliation{$^{24}$Rutgers University, New Brunswick, New Jersey 08854, USA}
\affiliation{$^{25}$Carnegie Mellon University, Pittsburgh, Pennsylvania 15213, USA}

\date{\today}

\begin{abstract}
  $^{2}\mathrm{H}(e,e'p)n$ cross sections have been measured at 4-momentum transfers of $Q^{2} = 4.5 \pm 0.5$ (GeV/c)$^{2}$
  over a range of neutron recoil momenta, $p_{\mathrm{r}}$,  reaching up to $\sim1.0$ GeV/c. The data were
  obtained at fixed neutron recoil angles $\theta_{nq} = 35^\circ$, $45^\circ$ and $75^{\circ}$  with respect to the 3-momentum
  transfer $\vec q$. The new data agree well with previous data which reached $p_{\mathrm{r}}\sim500$ MeV/c. At $\theta_{nq} = 35^\circ$
  and $45^\circ$, final state interactions (FSI), meson exchange currents (MEC) and isobar currents (IC) are suppressed and
  the plane wave impulse approximation (PWIA) provides the dominant cross section contribution. The new data are compared to recent
  theoretical calculations, where we observe a significant discrepancy for recoil momenta $p_{\mathrm{r}}>700$ MeV/c. 
\end{abstract}

\maketitle

The deuteron is the only bound two-nucleon system and serves as an ideal framework to study the strong nuclear force at the sub-Fermi distance scale, a region which is currently
practically unexplored and not well understood. Understanding the high momentum structure of the proton-neutron ($pn$) system is highly important for nuclear physics due to the observed 
dominance of short-range correlations (SRC) in nuclei at nucleon momenta above the fermi momentum. This dominance has been well established by a series of recent experiments
carried out at Jefferson Lab (JLab) \cite{Subedi1476, PhysRevLett.113.022501,PhysRevLett.122.172502, Schmidt_2020} and Brookhaven National Lab (BNL) \cite{PhysRevLett.97.162504}. In these experiments missing momenta up to $\sim1$ GeV/c have been probed and missing momentum distributions
have been compared to the high momentum part of theoretical deuteron momentum distributions.
Missing momenta up to $\sim1$ GeV/c have also been probed in a $^{3}\mathrm{He}(e,e'p)$ experiment \cite{PhysRevLett.94.082305,PhysRevLett.94.192302} but at a relatively low momentum transfer of 1.5 (GeV/c)$^{2}$ and at a kinematic region
(Bjorken $x_{\mathrm{Bj}}\sim1$) where the cross section is dominated by final state interactions. Due to FSI effects, the measurement of a certain missing momentum does not yet guarantee that the initial
bound nucleon with the same momentum is being measured.\\
\indent The most direct way to study the short range structure of the deuteron wave function
is via the exclusive deuteron electro-disintegration reaction at internal momenta $p_{\mathrm{r}}>300$ MeV/c. For $^{2}\mathrm{H}(e,e'p)n$, within the plane wave impulse approximation (PWIA),
the virtual photon couples to the bound proton which is subsequently ejected from the nucleus without further interaction with the recoiling system (neutron). The neutron carries a recoil
momentum, $p_{\mathrm{r}}$, equal in magnitude but opposite in direction to the initial state proton, $\vec{p}_{\mathrm{r}} \sim -\vec{p}_{\mathrm{i},p}$, thus providing information on the momentum
of the bound nucleon and its momentum distribution.\\
\indent In addition to the PWIA picture, the ejected nucleon undergoes final state interactions (FSI) with the recoiling nucleon. Other contributing processes are
the photon coupling to the exchanged mesons in the $pn$ system, generating meson exchange currents (MEC), or the photon exciting the bound nucleon into the
resonating state (mainly $\Delta$-isobar) with subsequent $\Delta N \to NN$ rescattering, referred to as isobar currents (IC). FSI, MEC and IC can significantly alter the recoiling neutron
momentum thereby obscuring the original momentum of the bound nucleon and reducing the possibility of directly probing the deuteron momentum distribution. \\
\indent Theoretically, MEC and IC are expected to be suppressed at $Q^{2}>1$ (GeV/c)$^{2})$ and Bjorken $x_{\mathrm{Bj}}\equiv Q^{2}/2M_{p}\omega>1$, where $M_{p}$ and $\omega$ are the proton mass and photon energy transfer, respectively \cite{sargsian_2015}.
The suppression of MEC can be understood from the fact that the estimated MEC scattering amplitude is proportional to  $(1 + Q^{2}/m^{2}_{\mathrm{meson}})^{-2}(1+Q^{2}/\Lambda^{2})^{-2}$, where $m_{\mathrm{meson}}\approx0.71$ GeV/c$^{2}$ and
$\Lambda^{2}\sim 0.8-1 $ (GeV/c)$^{2}$ \cite{Sargsian_2001}, this results in an additional $1/Q^{4}$ suppression as compared to the quasi-elastic contribution. Note that other meson exchange contributions that take place before the virtual photon interaction are included in the definition of the ground state wave function of the deuteron. IC can be suppressed kinematically by selecting $x_{\mathrm{Bj}}>1$, where one probes the lower energy ($\omega$) part of the deuteron quasi-elastic peak which is maximally away from the inelastic resonance
electro-production threshold. Previous deuteron electro-disintegration experiments performed at lower $Q^{2}$ ($Q^{2}<1$ (GeV/c)$^{2}$) (see Section 5 of Ref. \cite{sargsian_2015}) have helped quantify the contributions
from FSI, MEC and IC to the $^{2}\mathrm{H}(e,e'p)n$ cross sections and to determine the kinematics at which they are either suppressed (MEC and IC) or under control (FSI).  \\
\indent At large $Q^{2}$, FSI can be described by the Generalized Eikonal Approximation (GEA) \cite{Sargsian_2001,PhysRevC.56.1124,sargsian_2015}, which predicts a strong dependence of FSI on neutron recoil angles $\theta_{nq}$ (relative angles between recoil momenta $\vec{p}_{\mathrm{r}}$ and 3-momentum transfers $\vec{q}$).
GEA predicts FSI to be maximal for $\theta_{nq}\sim70^{\circ}$. This strong angular dependence has been found to lead to the cancellation of FSI at neutron recoil angles around $\theta_{nq}\sim40^{\circ}$ and $\theta_{nq}\sim120^{\circ}$. Because at
$\theta_{nq}\sim120^{\circ}$ ($x_{\mathrm{Bj}}<1$) IC are not negligible, the $x_{\mathrm{Bj}}>1$ ($\theta_{nq}\sim40^{\circ}$) kinematics are the preferred choice to suppress IC as well as FSI. \\
\indent The first $^{2}\mathrm{H}(e,e'p)n$ experiments at high $Q^{2}$ ($>1$ (GeV/c)$^{2}$) were carried out at JLab in Halls A \cite{PhysRevLett.107.262501} and B \cite{PhysRevLett.98.262502}. Both
experiments determined that the cross sections for fixed recoil momenta indeed exhibited a strong angular dependence with $\theta_{nq}$, peaking
at $\theta_{nq} \sim 70^{\circ}$ in agreement with GEA \cite{Sargsian_2001,PhysRevC.56.1124} calculations. In Hall B, the CEBAF Large Acceptance Spectrometer (CLAS) measured angular
distributions for a range of $Q^2$ values as well as momentum distributions. However, statistical limitations made it necessary to integrate over a wide angular range to determine momentum distributions
which are therefore dominated by  FSI, MEC and IC for $p_{\mathrm{r}}$ above $\sim 300$ MeV/c. \\
\indent In the Hall A experiment \cite{PhysRevLett.107.262501}, the pair of high resolution spectrometers (HRS) made it possible to measure the $p_{\mathrm{r}}$ dependence of the cross section for fixed $\theta_{nq}$
reaching recoil momenta up to $p_{\mathrm{r}}=550$ MeV/c at $Q^{2}=3.5\pm0.25$ (GeV/c)$^{2}$. For the first time, very different momentum distributions were found for $\theta_{nq}=35\pm5^{\circ}$
and $45\pm5^{\circ}$ compared to  $\theta_{nq}=75\pm5^{\circ}$. Theoretical models attributed this difference  to the suppression of FSI at the smaller angles ($\theta_{nq}=35, 45^{\circ}$) compared to FSI
dominance at $\theta_{nq}=75^{\circ}$ \cite{PhysRevLett.107.262501}. \\
\indent The experiment presented in this Letter takes advantage of the kinematic window previously found in the Hall A experiment \cite{PhysRevLett.107.262501} and extends the $^{2}\mathrm{H}(e,e'p)n$ cross section measurements
to $Q^{2}=4.5\pm0.5$ (GeV/c)$^{2}$ and recoil momenta up to $p_{\mathrm{r}}\sim 1$ GeV/c, which is almost double the maximum recoil momentum measured in Hall A \cite{PhysRevLett.107.262501}.
Measurements at such large $Q^{2}$ and high $p_{\mathrm{r}}$ required scattered electrons to be detected at $\sim 8.5$ GeV/c, which was only made possible with the newly commissioned Hall C Super High Momentum Spectrometer (SHMS).
At the selected kinematic settings with $35^{\circ} \leq \theta_{nq} \leq45^{\circ}$, MEC and IC are suppressed and FSI are under control giving access to high momentum components of the deuteron wave function.\\
\indent A 10.6 GeV electron beam was incident on a 10-cm-long liquid deuterium target (LD$_{2}$). The scattered electron and knocked-out proton were detected in coincidence
by the new SHMS and the existing High Momentum Spectrometer (HMS), respectively. The beam currents delivered by the accelerator ranged between 45-60 $\mu$A and the beam was rastered over a 2x2 mm$^{2}$ area to reduce the effects of localized boiling on the cryogenic targets.\\ 
\indent Both Hall C spectrometers have similar standard detector packages \cite{cyero_phdthesis}, each with four scintillator planes used for triggering, a pair of drift chambers used for tracking, and a calorimeter and gas \u{C}erenkov used for electron identification.
For each spectrometer, a logic signal was created from  the coincidence of hits in at least three of the four scintillator planes. The event trigger was the coincidence of these two signals. \\
\indent We measured three central recoil momentum settings: $p_{\mathrm{r}}=80, 580$ and $750$ MeV/c. At each of these settings, the electron arm (SHMS) was fixed and the proton arm (HMS) was rotated from smaller to larger angles corresponding to
the lower and higher recoil momentum settings, respectively. At these kinematic settings, the 3-momentum transfer covered a range of $2.4\lesssim|\vec{q}|\lesssim3.2$ GeV/c, which is more than twice the highest neutron recoil momentum
measured in this experiment. As a result, most of the virtual photon momentum is transferred to the proton, which scatters at angles relative to $\vec{q}$ in the range $0.4^{\circ}\lesssim \theta_{pq}\lesssim21.4^{\circ}$.
At these forward angles and large momenta transferred to the proton, the  process where the neutron is struck by the virtual photon is suppressed.\\
\indent Hydrogen elastic $^{1}\mathrm{H}(e,e'p)$ data were also taken at kinematics close to the deuteron $p_{\mathrm{r}}=80$ MeV/c setting for cross-checks with the spectrometer acceptance model using the Hall C Monte Carlo
simulation program, SIMC \cite{PhysRevC.64.054610}. Additional $^{1}\mathrm{H}(e,e'p)$ data were also taken at three other kinematic settings that covered the SHMS momentum acceptance range for the deuteron and were used for spectrometer optics optimization, 
momentum calibration and the determination of the spectrometer offsets and kinematic uncertainties \cite{cyero_phdthesis}.\\
\indent Identical event selection criteria were used for the hydrogen and deuteron data. The criteria were determined by making standard cuts on the spectrometer momentum fraction ($\delta$) to select a region for which the reconstruction optics
are well known,  a cut to restrict the HMS solid angle acceptance to events that passed directly through the collimator and not by re-scattering from the collimator edges, a reconstructed binding
energy cut (peak $\sim$ 2.22 MeV for the deuteron) to select true $^{2}\mathrm{H}(e,e'p)n$ coincidences, a coincidence time cut to select true coincidence events, a particle identification (PID) cut on the
SHMS calorimeter normalized total track energy to select electrons and not other sources of background (mostly pions), and a cut on the reconstructed HMS and SHMS reaction vertices to select events that 
originated from the same reaction vertex at the target (see online Supplemental Material). \\
\indent The experimental data yields for both hydrogen and deuteron data were normalized by the total charge and corrected for various inefficiencies. For $^{2}\mathrm{H}(e,e'p)n$, the corrections
were as follows\cite{cyero_phdthesis}: tracking efficiencies ($98.9 \%$-HMS, $96.4 \%$-SHMS), total live time ($92.3 \%$), proton loss inefficiency due to nuclear interactions in the HMS ($4.7 \%$) and
target boiling inefficiency ($4.2 \%$). The values in parentheses were averaged over all recoil momentum settings. \\
\indent For $^{1}\mathrm{H}(e,e'p)$, the corrected data yield was compared to SIMC calculations using Arrington's proton form factor (FF) parametrization \cite{PhysRevC.69.022201} to check the spectrometer acceptance
model. The ratio of data to simulation yield was determined to be $97.6\pm0.3 \%$ (statistical uncertainty only).\\
\indent The systematic uncertainties on the measured cross sections were determined from 
normalization and kinematic uncertainties in the beam energy and spectrometer angle/momentum settings. The individual
contributions from normalization uncertainties for each setting were determined to be (on average)\cite{cyero_phdthesis}: tracking efficiencies ($0.40 \%$-HMS, $0.59 \%$-SHMS),
and target boiling ($0.38 \%$) which were added in quadrature and determined to be about 0.81 $\%$ per setting. This result was then added quadratically to
the systematic uncertainties due to proton loss in HMS ($0.49 \%$), total live time ($3.0 \%$), total charge ($2.0\%$), target wall contributions ($\leq2.9\%$) and spectrometer acceptance ($1.4\%$),  which were the same for every setting,
to define the overall normalization uncertainty ($\leq5.3\%$).\\
\indent The systematic uncertainties due to the systematic error on the absolute beam energy and spectrometer angle/momentum settings were
determined point-to-point in ($\theta_{nq}$, $p_{\mathrm{r}}$) bins for each recoil momentum setting, and added in quadrature for overlapping $p_{\mathrm{r}}$ bins. 
For $\theta_{nq}= 35^{\circ}, 45^{\circ}$ and $75^{\circ}$ (presented in this Letter) the overall kinematic uncertainty varied below 6.5$\%$.
The total uncertainty was defined as the quadrature sum of the normalization ($\leq$ 5.3$\%$), kinematic ($\leq$ 6.5$\%$) and statistical ($\sim20-30\%$ on average) uncertainties.\\
\begin{figure*}[t]
\centering        
\includegraphics[scale=0.45]{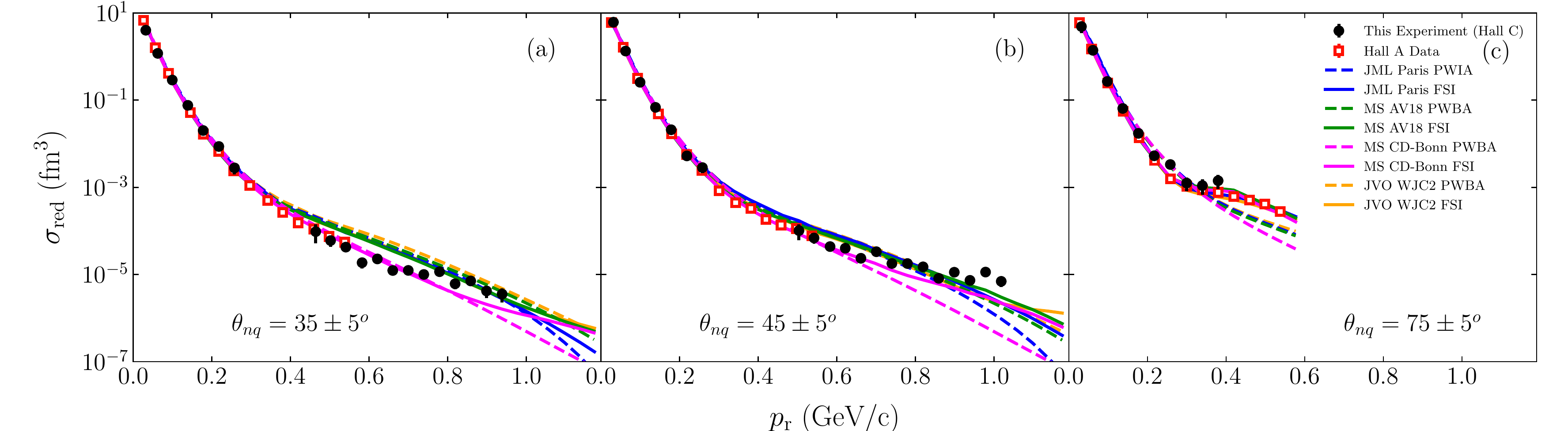}
\caption{(Color online). The reduced cross sections $\sigma_{\mathrm{red}}(p_{\mathrm{r}})$ as a function of neutron recoil momentum $p_{\mathrm{r}}$ are shown in (a)-(c) for recoil angles $\theta_{nq}=35^{\circ}, 45^{\circ}$ and $75^{\circ}$, respectively,
with a bin width of $\pm 5^{\circ}$. The data are compared to the previous Hall A experiment (red square) results \cite{PhysRevLett.107.262501} as well as the theoretical reduced cross sections using the Paris (blue),
AV18 (green), CD-Bonn (magenta) and WJC2 (orange) $NN$ potentials.}
\label{fig:fig1}
\end{figure*}
\indent The data were radiatively corrected for each bin in ($\theta_{nq}, p_{\mathrm{r}}$) by multiplying the measured cross sections by the ratio of the calculated particle yield excluding and including radiative effects.
The SIMC simulation code was used for these calculations with the Deuteron Model by Laget including FSI \cite{LAGET2005}.
For each bin in ($\theta_{nq}, p_{\mathrm{r}}$), the averaged $^{2}\mathrm{H}(e,e'p)n$ kinematics was calculated and used in the bin centering correction factor defined as:
$f_{\mathrm{bc}} \equiv \sigma_{\mathrm{avg.kin}} / \bar{\sigma}$, where $\sigma_{\mathrm{avg.kin}}$ is the cross section calculated at the averaged kinematics and $\bar{\sigma}$ is the cross section averaged
over the kinematic bin. The systematic uncertainties associated with the radiative and bin-centering corrections were investigated using the Laget PWIA and FSI models but negligible effects
on the cross sections were found (see online Supplemental Material).
The experimental and theoretical reduced cross sections were extracted and are defined as follows:
\begin{equation}
\sigma_{\mathrm{red}} \equiv \frac{\sigma_{\mathrm{exp(th)}}}{E_{\mathrm{f}}p_{\mathrm{f}}f_{\mathrm{rec}}\sigma_{\mathrm{cc1}}},
\label{eq:1}
\end{equation}
\noindent where $\sigma_{\mathrm{exp(th)}}$ is the 5-fold experimental (or theoretical) differential cross section $\frac{d^{5}\sigma}{d\omega d\Omega_{e} d\Omega_{p}}$,
($E_{\mathrm{f}}, p_{\mathrm{f}}$) are the final proton energy and momentum, respectively, $f_{\mathrm{rec}}$ 
is a recoil factor \cite{cyero_phdthesis} obtained by integrating over the binding energy of the bound state in the 6-fold differential cross section and $\sigma_{\mathrm{cc1}}$ is the
de Forest \cite{DEFOREST1983} electron-proton off-shell cross section calculated using the FF parametrization of Ref. \cite{PhysRevC.69.022201}.
Within the PWIA, $\sigma_{\mathrm{red}}$ corresponds to the PWIA cross section from the scattering of a proton in the deuteron. \\
\indent Figure \ref{fig:fig1} shows the extracted experimental and theoretical reduced cross sections as a function of $p_{\mathrm{r}}$ for three recoil angle settings at $Q^{2}=4.5\pm0.5$ (GeV/c)$^{2}$.
For the two highest momentum settings ($p_{\mathrm{r}}=580, 750$ MeV/c), a weighted average of the reduced cross sections were taken in the overlapping regions of $p_{\mathrm{r}}$. The results from the previous experiment \cite{PhysRevLett.107.262501} at a $Q^{2}=3.5\pm0.25$ (GeV/c)$^{2}$ are plotted as well (red square).
The data are compared to theoretical calculations using wave functions determined from the charge-dependent Bonn (CD-Bonn) \cite{PhysRevC.63.024001}, Argonne $v_{18}$ (AV18) \cite{PhysRevC.51.38}, Paris \cite{PhysRevC.21.861} and WJC2 \cite{Gross_2007} $NN$ potentials. The theoretical calculations
for the CD-Bonn (magenta) and AV18 (green) potentials were performed by Sargsian \cite{PhysRevC.82.014612} within the GEA, referred to as MS, and those for the Paris potential (blue) were by Laget \cite{LAGET2005} within the diagrammatic approach, referred to as JML.
For the WJC2 (orange) potential, the calculations were carried out by Ford, Jeschonnek and Van Orden \cite{PhysRevC.90.064006} using a Bethe-Salpeter-like formalism for two-body bound states, which will be labeled JVO.
The calculations use different FF parametrizations which can lead to a $\sim$5.8 -6.6$\%$ variation of the theortical cross section.\\
\indent The difference between the deuteron wave functions with CD-Bonn, Paris, AV18 and WJC2 potentials is 
how the $NN$ potential is modeled based on the empirical $NN$ scattering data.
The CD-Bonn model is based on the One-Boson-Exchange Potential (OBEP) approach in which the 
nucleon-meson-meson couplings are constrained to describe the $NN$ scattering phase shifts
extracted from the data. The interaction potential represents the static limit of 
this potential. In contrast, the WJC2 is a OBEP derived within the Covariant Spectator Theory (CST) \cite{PhysRev.186.1448, PhysRevD.10.223, PhysRevC.26.2203, PhysRevC.26.2226}
which requires comparatively few parameters while still producing a high-precision fit to the $NN$ scattering data.
The Paris and AV18 are purely phenomenological potentials where a 
Yukawa type interaction is introduced and parameters are fitted to describe the 
same $NN$ scattering phase-shifts. The major difference between the CD-Bonn and Paris/AV18/WJC2 
potentials is that the former predicts a much softer repulsive interaction at short distance which 
results in a smaller high momentum component in the deuteron wave function in momentum space.
The effects of these local approximations on the $NN$ potential are shown in Fig. 2 of Ref. \cite{PhysRevC.63.024001}.\\
\indent For all recoil angles shown in Fig. \hyperref[fig:fig1]{1} at recoil momenta $p_{\mathrm{r}}\leq250$ MeV/c, the cross sections are well reproduced by all models when FSI are included.
The agreement at $p_{\mathrm{r}}\leq250$ MeV/c can be understood from the fact that this region corresponds to the long-range part of the $NN$ potential where the One Pion Exchange Potential (OPEP)
is well known and common to all modern potentials. \\
\indent Beyond $p_{\mathrm{r}}\sim250$ MeV/c at $\theta_{nq}=35^{\circ}$ and $45^{\circ}$ (Figs. \hyperref[fig:fig1]{1(a), 1(b)}), the JML,
MS AV18 and JVO models increasingly differ from the MS CD-Bonn calculation. In this region, the JML and MS AV18 cross sections are dominated by the PWIA and in good agreement up to $p_{\mathrm{r}}\sim700$ MeV/c whereas
the JVO PWIA falls off with a comparatively smaller cross section at $\theta_{nq}=35^{\circ}$. The MS CD-Bonn cross sections in contrast are generally smaller than the JML, MS AV18 and JVO in this region.
In addition, for $\theta_{nq}=35^{\circ}$, they are dominated by the PWIA up to $p_{\mathrm{r}}\sim800$ MeV/c
(Fig. \hyperref[fig:fig1]{1(a)}) while for $\theta_{nq}=45^{\circ}$ FSI start to contribute already above 600 MeV/c (Fig. \hyperref[fig:fig1]{1(b)}).\\
\indent For recoil momenta $p_{\mathrm{r}} \sim 0.55-1.0$ GeV/c (Figs. \hyperref[fig:fig1]{1(a), 1(b)}), all models exhibit a
steeper fall-off compared to data. This discrepancy was quantified by doing a linear fit to the data and each of the PWIA calculations. A difference of at least
4.2 standard deviations was found between the data and theory slopes which corresponds to a probability $\leq 1.1\times 10^{-5}$ (very unlikely) that the observed discrepancy is due to a statistical fluctuation. \\ 
\indent At $\theta_{nq}=75^{\circ}$ (Fig. \hyperref[fig:fig1]{1(c)}) and $p_{\mathrm{r}}>180$ MeV/c, FSI become the dominant contribution to the cross sections for all models which exhibit a similar
behavior (smaller fall-off) that overshadows any possibility of extracting the approximate momentum distributions.\\
\indent To quantify the discrepancy observed between data and theory in Fig. \ref{fig:fig1}, the ratio of the experimental and theoretical reduced cross sections ($\sigma_{\mathrm{red}}$) to the
deuteron momentum distribution calculated using the CD-Bonn potential ($\sigma^{\text{CD-Bonn PWIA}}_{\mathrm{red}}$) \cite{PhysRevC.63.024001} is shown in Fig. \ref{fig:fig2}. \\
\indent For $\theta_{nq}=35^{\circ}$ and $45^{\circ}$ (Figs. \hyperref[fig:fig2]{2(a),(b)}), the data are best described by the MS CD-Bonn PWIA calculation for recoil momenta up
to $p_{\mathrm{r}}\sim700$ MeV/c and $\sim600$ MeV/c, respectively. Furthermore, the agreement between the Halls A and C data validates the Hall A approach of selecting a kinematic
region where recoil angles are small and FSI are reduced. \\
\indent At larger recoil momenta, where the ratio $R>1$ and increasing with $p_{\mathrm{r}}$, for $\theta_{nq}=35^{\circ}$ FSI start to dominate at
$p_{\mathrm{r}} \gtrsim 800$ MeV/c for the MS CD-Bonn
\begin{figure}[!t]
\includegraphics[scale=0.5]{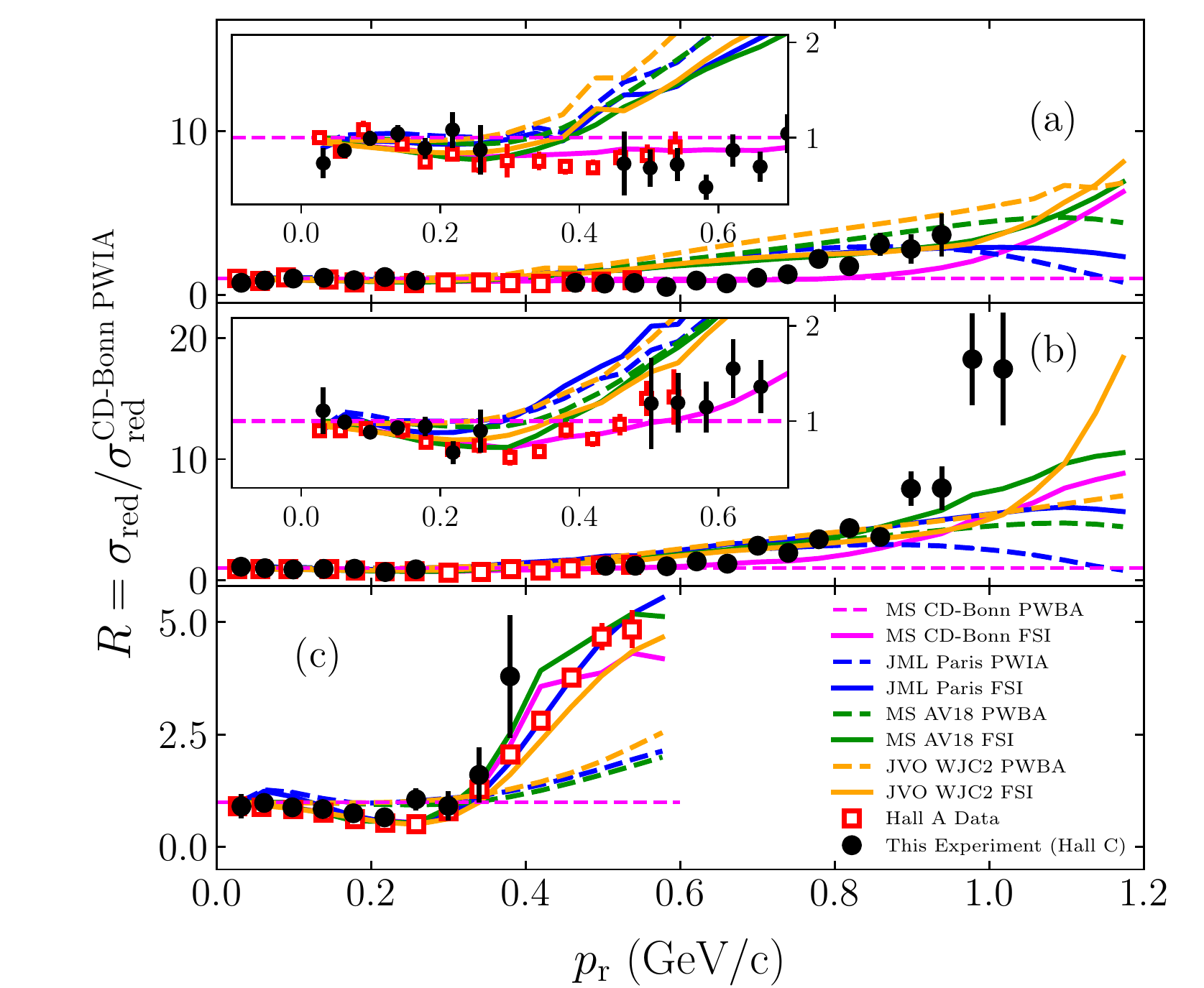}
\caption{(Color online). The ratio $R(p_{\mathrm{r}}$) is shown in (a)-(c) for $\theta_{nq}=35^{\circ}, 45^{\circ}$ and $75^{\circ}$, respectively, each with a bin width of $\pm 5^{\circ}$.
  The dashed reference (magenta) line refers to MS CD-Bonn PWIA calculation (or momentum distribution) by which the data and all models are divided.
  Insets: Blowup of the subfigures for $p_{\mathrm{r}}\leq$0.7 GeV/c.}
\label{fig:fig2}
\end{figure}
calculation while the other models predict still relatively small FSI below 900 MeV/c.
At $\theta_{nq}=45^{\circ}$, the FSI dominance starts earlier for all models above 800 MeV/c and for the MS CD-Bonn based calculation above 600 MeV/c. \\
\indent Overall, it is interesting to note that none of the calculations can reproduce the measured $p_{\mathrm{r}}$ dependence above 600 MeV/c in a
region where FSI are still relatively small $(<30\%)$. This behavior of the data is new and additional data in this kinematic region are necessary
to improve the statistics. \\
\indent At $\theta_{nq}=75^{\circ}$ (Fig. \hyperref[fig:fig2]{2(c)}), FSI are small below $p_{\mathrm{r}}\sim180$ MeV/c, but do not exactly cancel the PWIA/FSI interference term in the scattering amplitude which results in a small dip in this region in agreement with the data.
At $p_{\mathrm{r}}>300$ MeV/c ($\theta_{nq}=75^{\circ}$), the data were statistically limited as our focus was on the smaller recoil angles. The Hall A data, however, show a reasonable agreement with the FSI from all models
which gives us confidence in our understanding of FSI at the smaller recoil angles. \\
\indent To summarize, this experiment extended the previous Hall A cross section measurements on the $^{2}\mathrm{H}(e,e'p)n$ reaction to 
$p_{\mathrm{r}}>500$ MeV/c at kinematics where FSI were expected to be small and the cross sections were dominated by PWIA and sensitive to the
short range part of the deuteron wave function. The experimental reduced cross sections were extracted and found to be in good agreement with the Hall A data at lower recoil momenta where they overlap.
Furthermore, the MS CD-Bonn model was found to be significantly different than the JML, MS AV18 or JVO models and was able to partially describe the data over a larger range in $p_{\mathrm{r}}$.
At the higher recoil momenta provided by this experiment ($p_{\mathrm{r}}>700$ MeV/c), however, all models were unable to describe the data,
potentially illustrating the limit to which a non-relativistic wave function from the solution to the Schr\"{o}dinger equation is valid and able to describe
experimental data that probe the high momentum region of the $np$ system in the most direct way possible.
The new data set is also ideal for testing fully relativistic deuteron models based on light-front \cite{Frankfurt:1981mk} or covariant \cite{Buck:1979ff} formalisms.
In this respect the current effective field theories (EFT) based models \cite{Reinert_EFT2018} are non-relativistic and might not have direct relevance to our data.
Additional measurements of the $^{2}\mathrm{H}(e,e'p)n$ would be required to reduce the statistical uncertainties in this very high recoil
momentum region ($p_{\mathrm{r}}>500$ MeV/c) to better understand the large deviations observed between the different models and data.\\
\indent We acknowledge the outstanding support of the staff of the Accelerator and Physics Divisions at Jefferson Lab
as well as the entire Hall C staff, technicians, graduate students and users who took shifts or contributed
to the equipment for the Hall C upgrade, making all four commissioning experiments possible. We would also like to
thank Misak Sargsian, J.M. Laget, Sabine Jeschonnek and J.W. Van Orden for providing the theoretical calculations as well as the useful
discussions we had on this topic. This work was supported in part by the U.S. Department of Energy (DOE), Office of Science, Office of Nuclear Physics
under grant No. DE-SC0013620 and contract DE-AC05-06OR23177, the Nuclear Regulatory Commission (NRC) Fellowship
under grant No. NRC-HQ-84-14-G-0040, the Doctoral Evidence Acquisition (DEA) Fellowship and the Natural Sciences and Engineering Research Council of Canada (NSERC).

\nocite{barlow2002systematic}
\nocite{barlow2017}

\bibliography{ms}
\end{document}


\title{Supplementary Materials: Probing the Deuteron at Very Large Internal Momenta}
\maketitle
\section{\large Event Selection Cuts}
\begin{figure}[!h]
\includegraphics[scale=0.33]{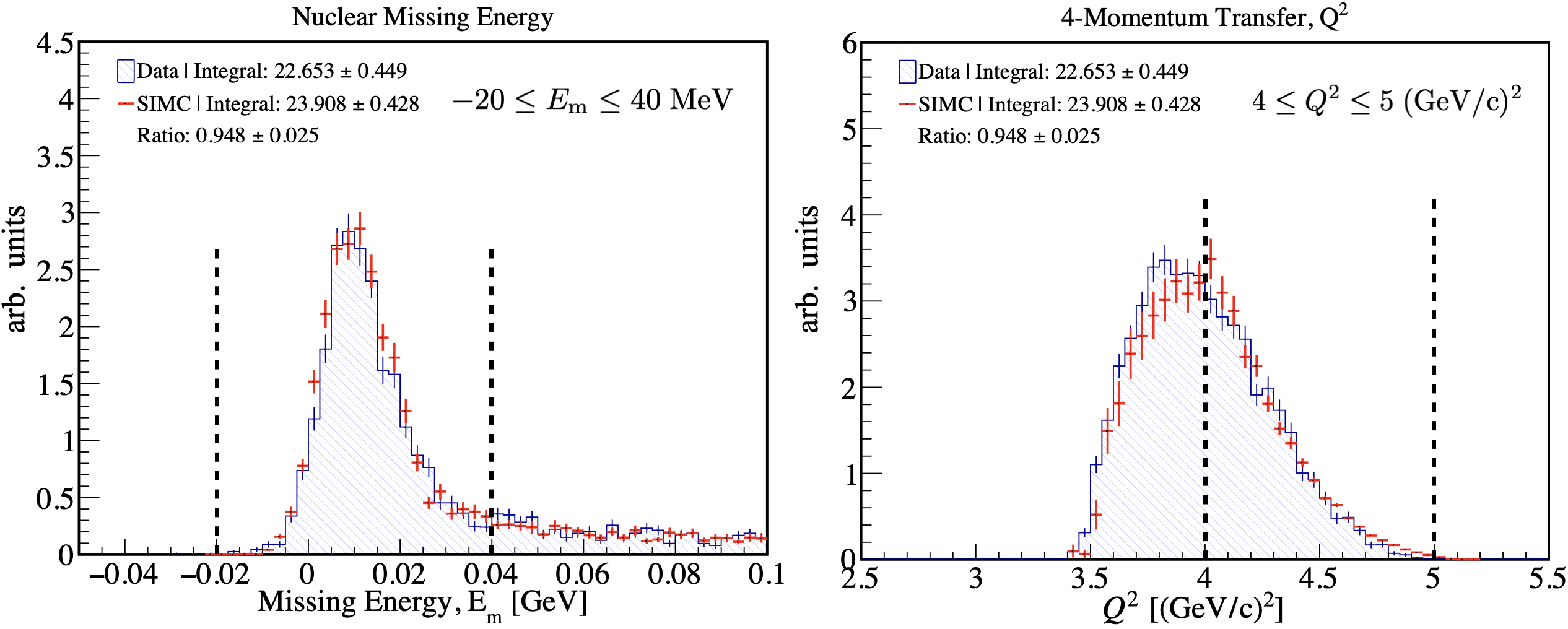}
\caption{Event selection cuts on missing energy (left) and 4-momentum transfer (right).}
\label{fig:Em_Q2_cuts}
\includegraphics[scale=0.33]{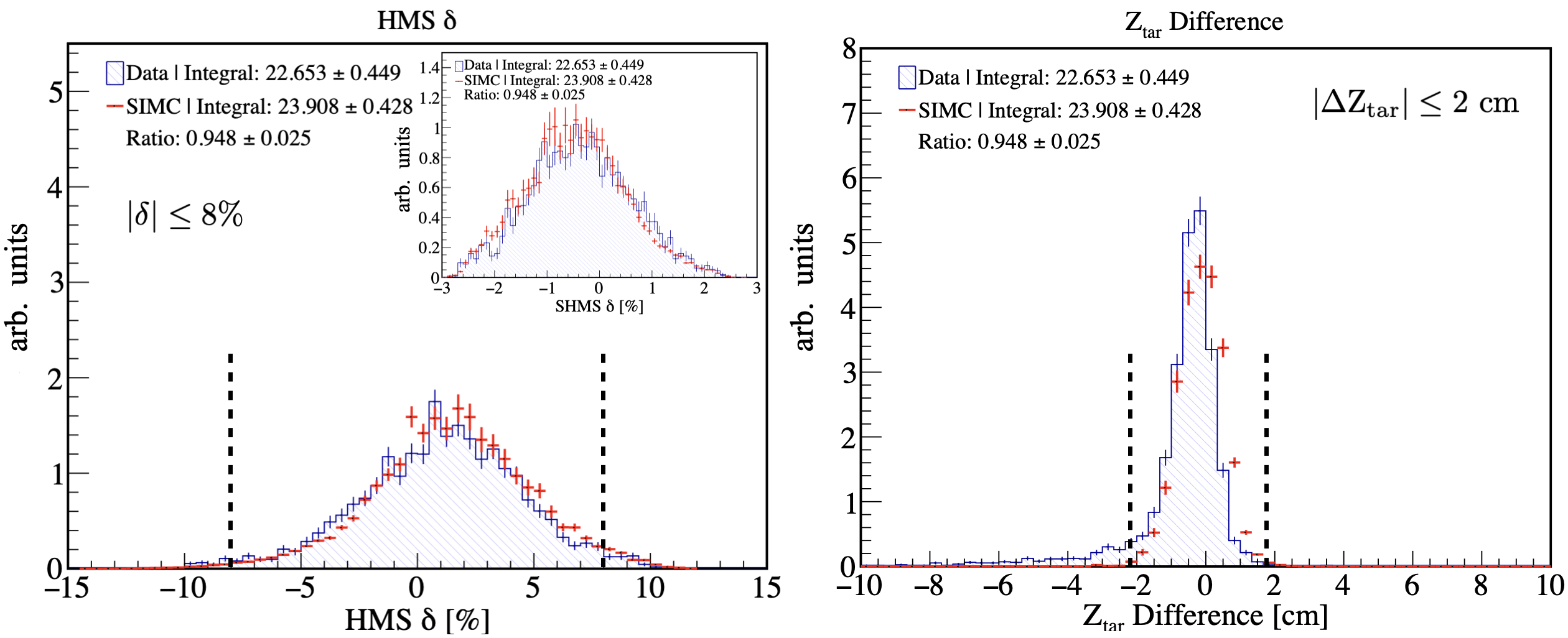}
\caption{Acceptance cut on HMS momentum fraction (left) and event selection cut on the difference between the $z$-reaction vertex on both spectrometers (right).
  Inset (left): The SHMS momentum acceptance range.}
\label{fig:delta_Ztar_cuts}
\end{figure}

\begin{figure}[!h]
\includegraphics[scale=0.23]{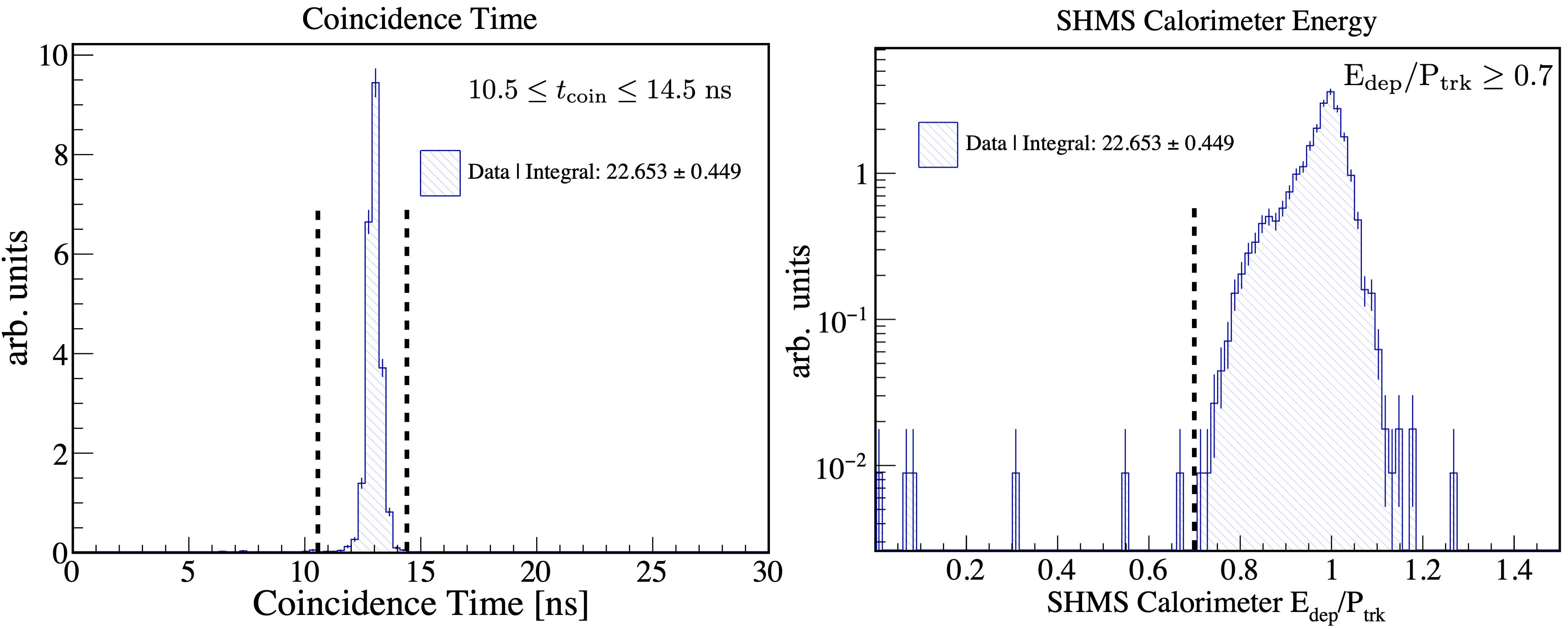}
\caption{Event selection cuts on the electron-proton ($ep$) coincidence time (left) and total deposited energy on calorimeted normalized by the particle track momentum (right).}
\label{fig:coin_ecal_cuts}
\end{figure}

\begin{figure}[!h]
\includegraphics[scale=0.24]{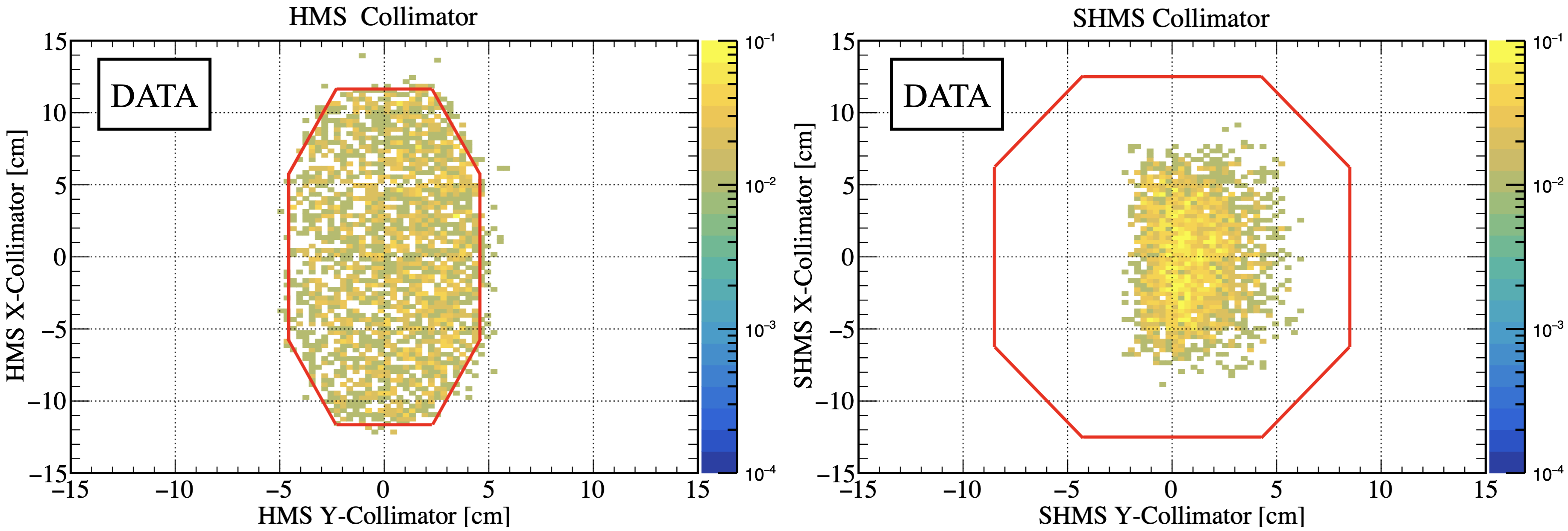}
\caption{(left) Geometrical acceptance cut on reconstructed events projected at the HMS collimator. (right) The SHMS events (correlated with HMS events on left plot),
  projected at the SHMS collimator. The $z$-axis (color palette) is in arbitrary units.}
\label{fig:data_coll_cuts}
\end{figure}

\begin{figure}[!h]
\includegraphics[scale=0.25]{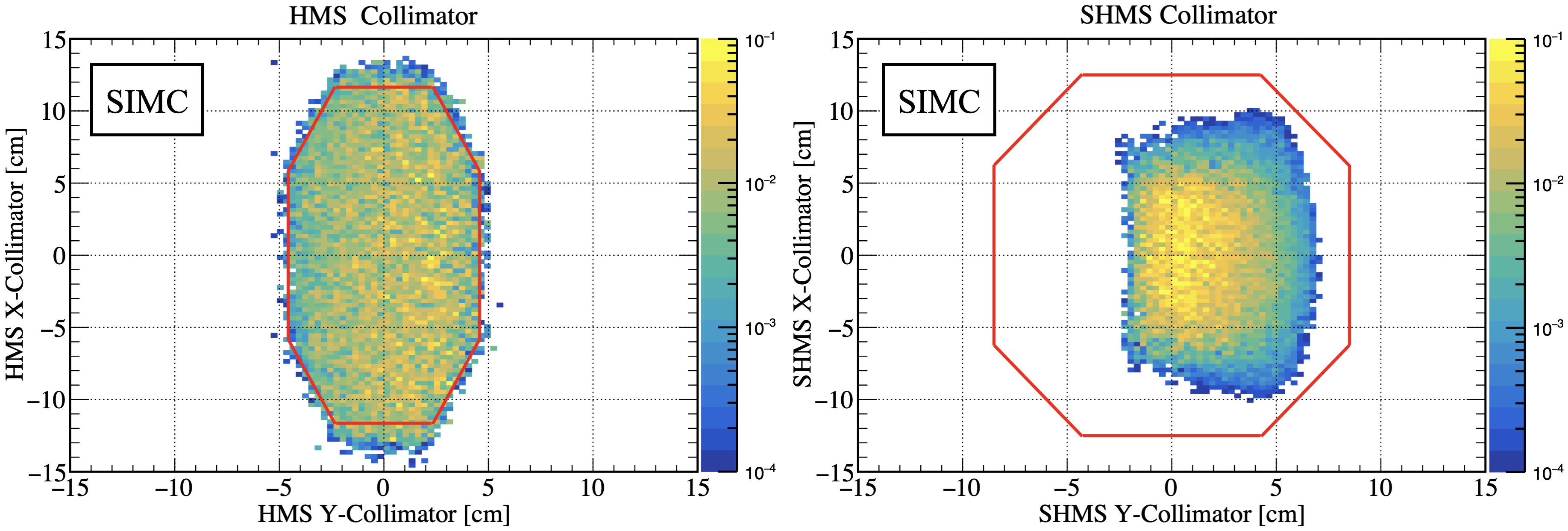}
\caption{Same as Fig. \ref{fig:data_coll_cuts}, but for simulation.  The $z$-axis (color palette) is in arbitrary units.}
\label{fig:simc_coll_cuts}
\end{figure}

\indent Figures \ref{fig:Em_Q2_cuts}-\ref{fig:simc_coll_cuts} show the event selection cuts for data (blue hatched) and SIMC (red data points) for the deuteron 80 MeV/c kinematic setting at $\theta_{nq}=35\pm5^{\circ}$.
The black dashed or red solid lines (for collimators) represent the cuts or geometrical (collimators) boundaries used to select true $^{2}$H$(e,e'p)n$ coincidence events.
The exact same cuts were also applied to the 580 and 750 MeV/c settings. Each histogram has all the other event selection cuts described except a self cut.
The integrated counts (within the cut boundary) for data and simulation as well as the data-to-simulation yield ratio is shown (top left) for each selection cut.
The data yield has been normalized by the total charge and corrected for the inefficiencies described in the Letter. The FSI model from J.M. Laget was used in the simulation for the plots shown below.
\\
\indent Figure \ref{fig:Em_Q2_cuts} (left) shows the primary cut used to select true $^{2}\mathrm{H}(e,e'p)n$ coincidence events by requiring a missing
energy cut around the deuteron binding energy ($\sim$2.2 MeV) using the missing energy formula, $E_{\mathrm{m}} = \omega - T_{p} - T_{\mathrm{r}}$, where
$T_{p}$ is the final proton kinetic energy and $T_{\mathrm{r}}$ is the recoil particle kinetic energy which is calculated from the electron and proton
4-momentum vectors assuming an exclusive three-body final state with a recoiling neutron. The peak is not exactly at the deuteron binding energy because
energy loss corrections have not been applied to the data nor SIMC. \\
\indent Figure \ref{fig:Em_Q2_cuts} (right) shows a
kinematical cut made on the 4-momentum transfer at $Q^{2} = 4.5\pm0.5$ (GeV/c)$^{2}$
to select events only at the highest possible momentum transfers to further suppress MEC and IC. \\
\indent Figure \ref{fig:delta_Ztar_cuts} (left) shows a momentum acceptance ($\delta$) cut made on the HMS in the range $-8\%\leq\delta\leq8\%$, where the optics reconstruction
is reliable and well understood from previous experiments in the 6 GeV era. The inset shows the SHMS momentum acceptance which was constrained by that of the HMS to be
$|\delta|\lesssim$3$\%$ which is well within the reliable SHMS momentum acceptance range of $-10 \leq \delta \leq22 \%$. \\
\indent Figure \ref{fig:delta_Ztar_cuts} (right) shows a reaction vertex cut
on the difference between the HMS and SHMS reconstructed reaction vertices along the $z$-coordinate (target length). The cut is made at $\pm2$ cm relative to the histogram peak.
If the events originated from the same reaction vertex (i.e., true coincidences), the difference should peak at zero with a finite resolution width. If the events are uncorrelated
(i.e., accidental coincidences), however, the reconstruction along the $z_{\mathrm{v}}$-vertex can vary significantly between the two spectrometers, which contribute to the tails of the
distribution.\\
\\
\indent Figure \ref{fig:coin_ecal_cuts} (left) shows a coincidence time cut on the $ep$ coincidence time spectrum formed by requiring a logic signal of
at least 3 out of 4 scintillator planes between both spectrometers. The spectrum is very clean as the typical 4 ns beam bunch structure due to accidental
coincidences is not observed. \\
\indent Figure \ref{fig:coin_ecal_cuts} (right) shows a particle identification cut on the SHMS calorimeter total energy
deposited normalized by the particle track momentum. This cut was used to separate electrons from background (pions), however, the spectrum shows a very clean distribution
with a peak at about one indicating the detected particles were electrons.
\clearpage
\indent Figure \ref{fig:data_coll_cuts} (left) shows a geometrical (red octogon) cut on the reconstructed events projected at the HMS collimator to ensure
that all events that enter the spectrometer pass through the collimator, and not re-scatter at the edges. This cut is needed since protons that enter the
HMS can punch through the collimator with a probability that is related to the proton-Tungsten total cross section.\\
\indent Figure \ref{fig:data_coll_cuts} (right)
shows the correlated events projected at the SHMS collimator which clearly shows that the events are well within the collimator which indicates that the SHMS
acceptance is driven by that of the HMS. Figure \ref{fig:simc_coll_cuts} shows the same acceptance cuts as previously described, but for simulation. The colored
$z$-axis in both figures is given in arbitrary units.\\
\section{\large Deuteron Kinematic Distributions }
\begin{figure}[!h]
\includegraphics[scale=0.17]{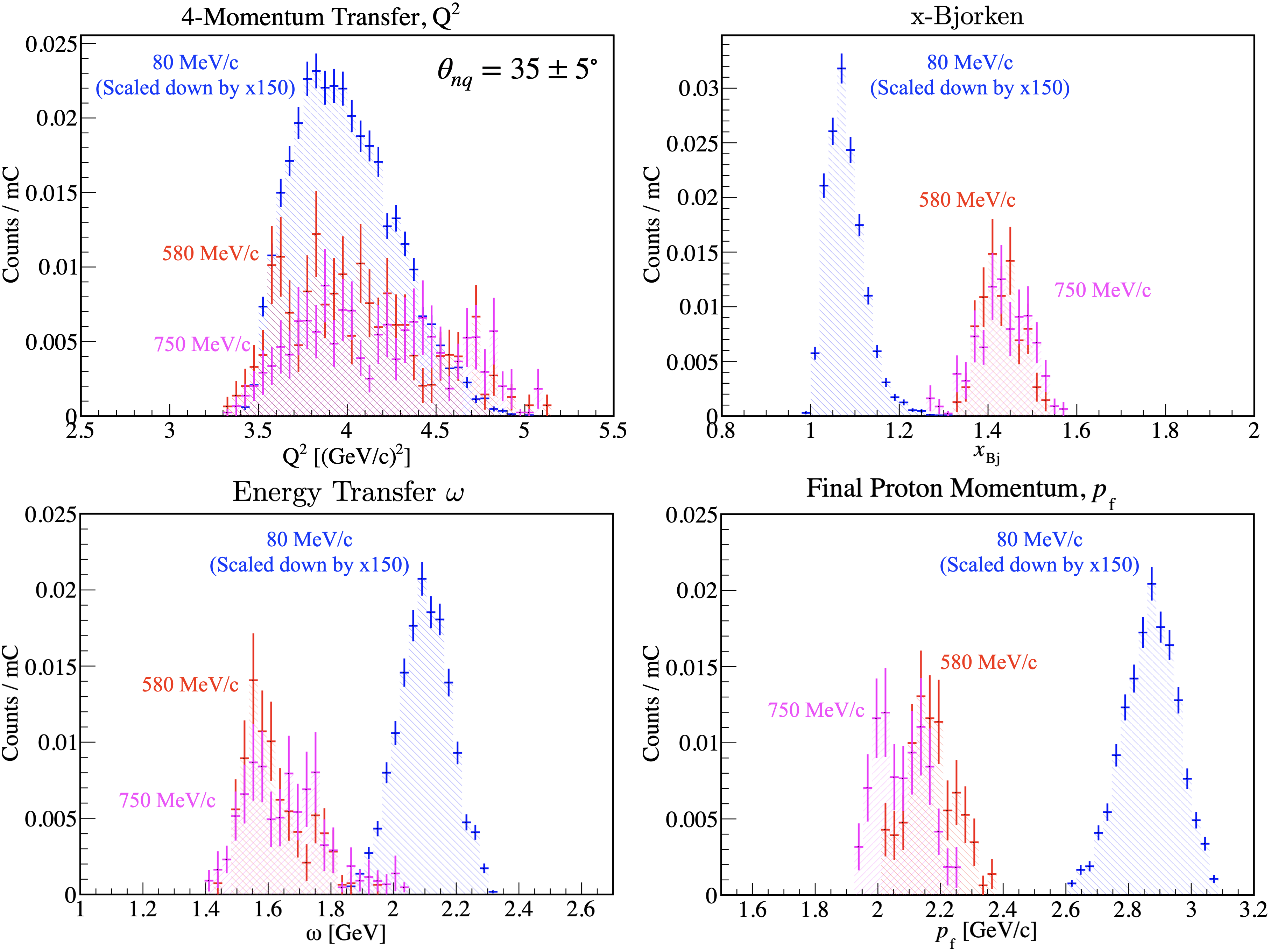}
\caption{Deuteron kinematic distributions for the 80 (blue), 580 (red) and 750 (magenta) MeV/c setting at $\theta_{nq}=35^{\circ}$.}
\label{fig:d2_kin_35deg}
\end{figure}

\begin{figure}[!h]
\includegraphics[scale=0.17]{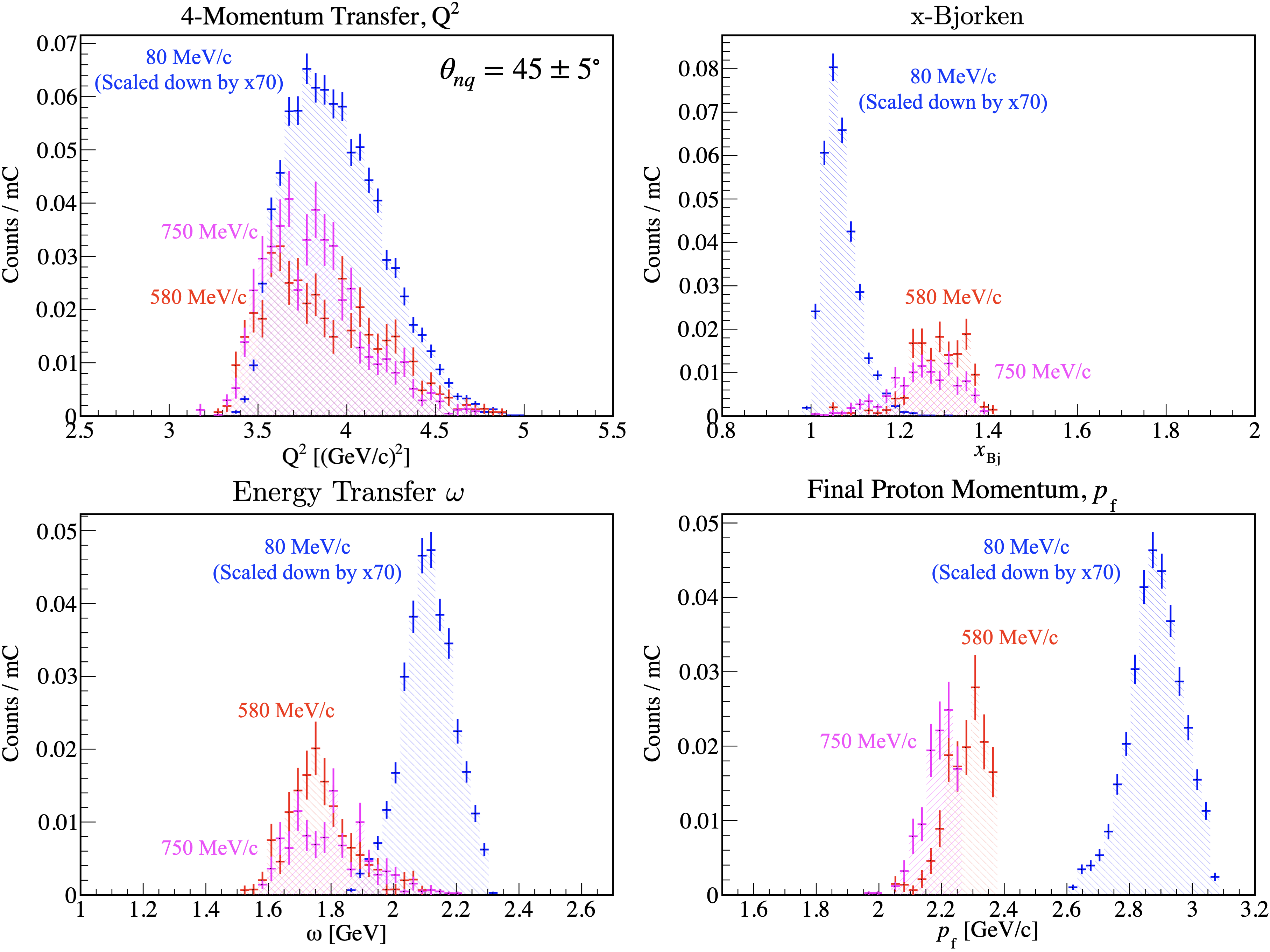}
\caption{Same as Fig. \ref{fig:d2_kin_35deg}, but for $\theta_{nq}=45^{\circ}$}
\label{fig:d2_kin_45deg}
\end{figure}

\indent Figures \ref{fig:d2_kin_35deg} and \ref{fig:d2_kin_45deg} show the fundamental kinematic variable distributions for each of the
central missing momentum settings at $\theta_{nq}=35^{\circ}$ and $45^{\circ}$. The data has been normalized by the total charge and corrected
for the inefficiencies described in the Letter. The same event selection cuts mentioned in the previous section were applied. 
\section{\large Estimate of the Target Cell Endcaps Contribution to the Data Yield}
\begin{figure}[!h]
\includegraphics[scale=0.33]{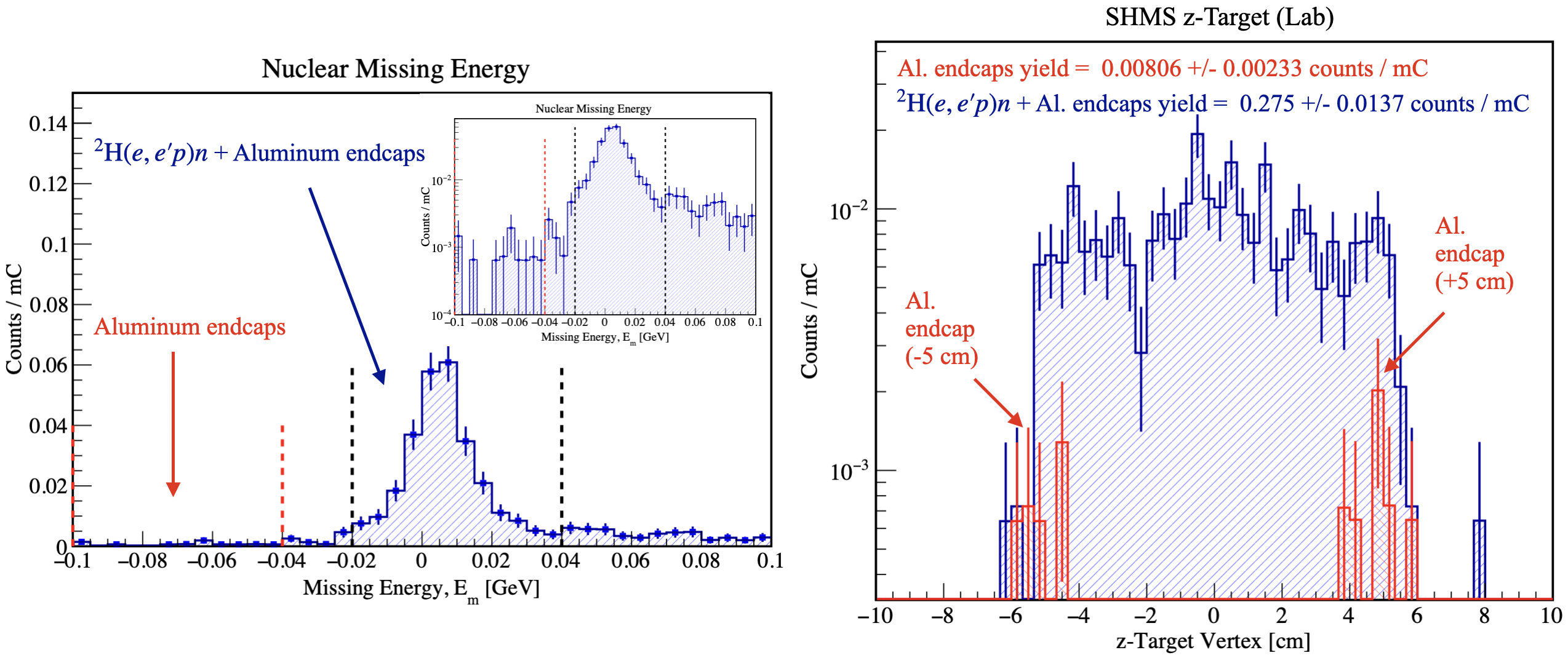}
\caption{(left) Missing energy spectrum for the deuteron 580 MeV/c setting with event selection region corresponding to Aluminum endcaps (in red) and deuteron missing energy peak over
  a 60 MeV range, each.  Inset (left): Missing energy spectrum on a logarithmic scale. (right) The SHMS $z$-reaction vertex corresponding to
  the specified region in the missing energy spectrum.}
\label{fig:tgt_wall}
\end{figure}

\indent To estimate the contribution to the yield due to the electron interaction with the aluminum endcaps of the target cell,
a sample of events were selected in the negative part of the missing energy spectrum using the deuteron high missing momentum settings (580 and 750 MeV/c).
We assume that the contribution due to the target endcaps is constant across the missing energy spectrum, therefore, by selecting a sample in the
negative part of the spectrum over a specific range, we can estimate the endcaps contribution beneath the deuteron missing energy peak over the same range.\\
\indent Figure \ref{fig:tgt_wall} shows the missing energy spectrum for the 580 MeV/c setting (left) and the corresponding reconstructed SHMS $z$-vertex (right) for the specified
range. The integral over endcaps and $^{2}\mathrm{H}(e,e'p)n$ events show that the contribution from the cell walls has an upper limit of 0.00806/0.275 $\sim$ 0.0293 or approximately 2.9 $\%$
as the integral was done over all recoil angles, $\theta_{nq}$. This result was added quadratically to the overall normalization error.
\section{\large Cross-Section Extraction}
\begin{figure}[!h]
\includegraphics[scale=0.30]{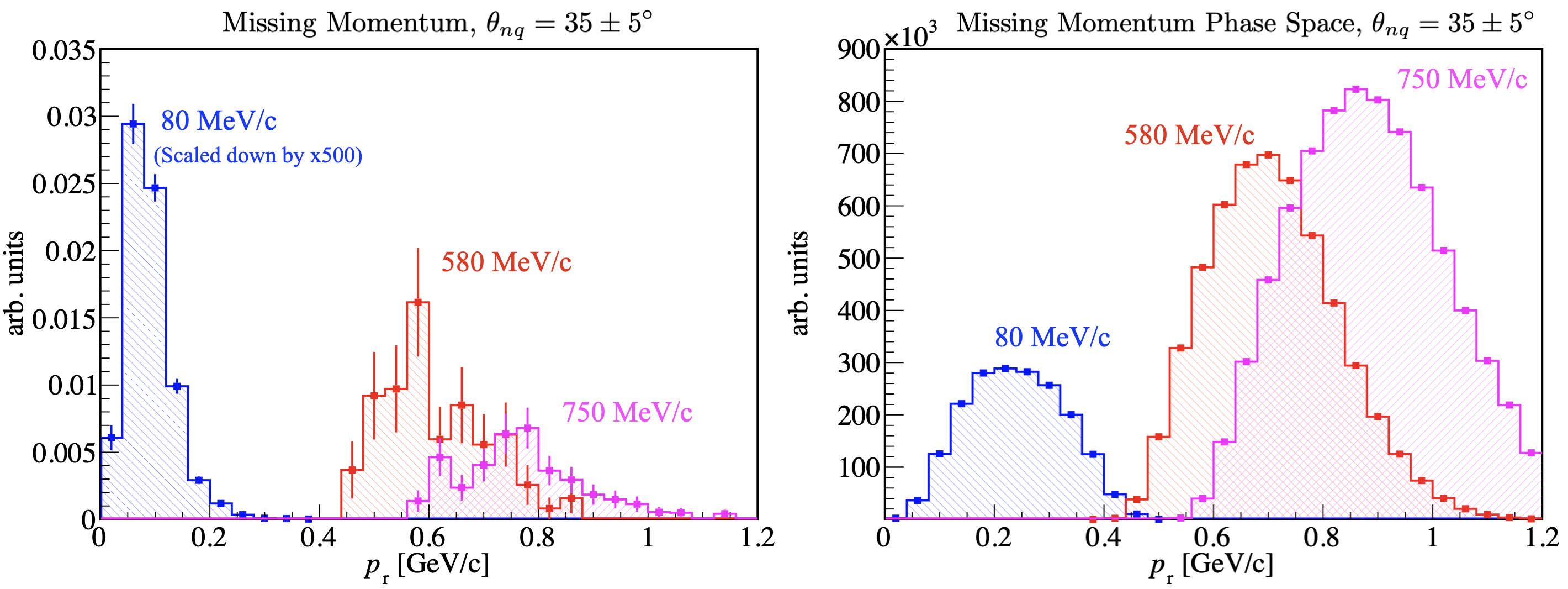}
\includegraphics[scale=0.30]{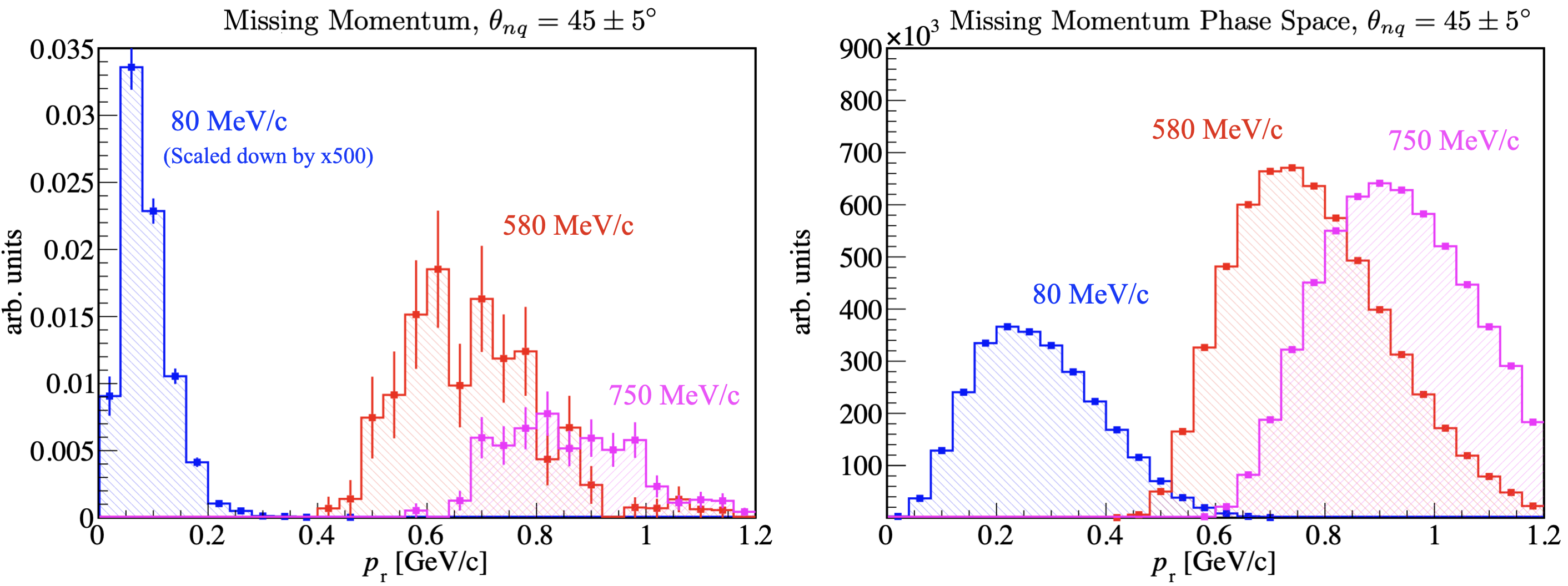}
\caption{(left column) Experimental neutron recoil (``missing'') momentum distribution for each of the three central settings. (right column) Monte Carlo (un-weighted) events generated over
  the spectrometers' phase space volume binned in missing momentum.}
\label{fig:Pm_Ps}
\includegraphics[scale=0.30]{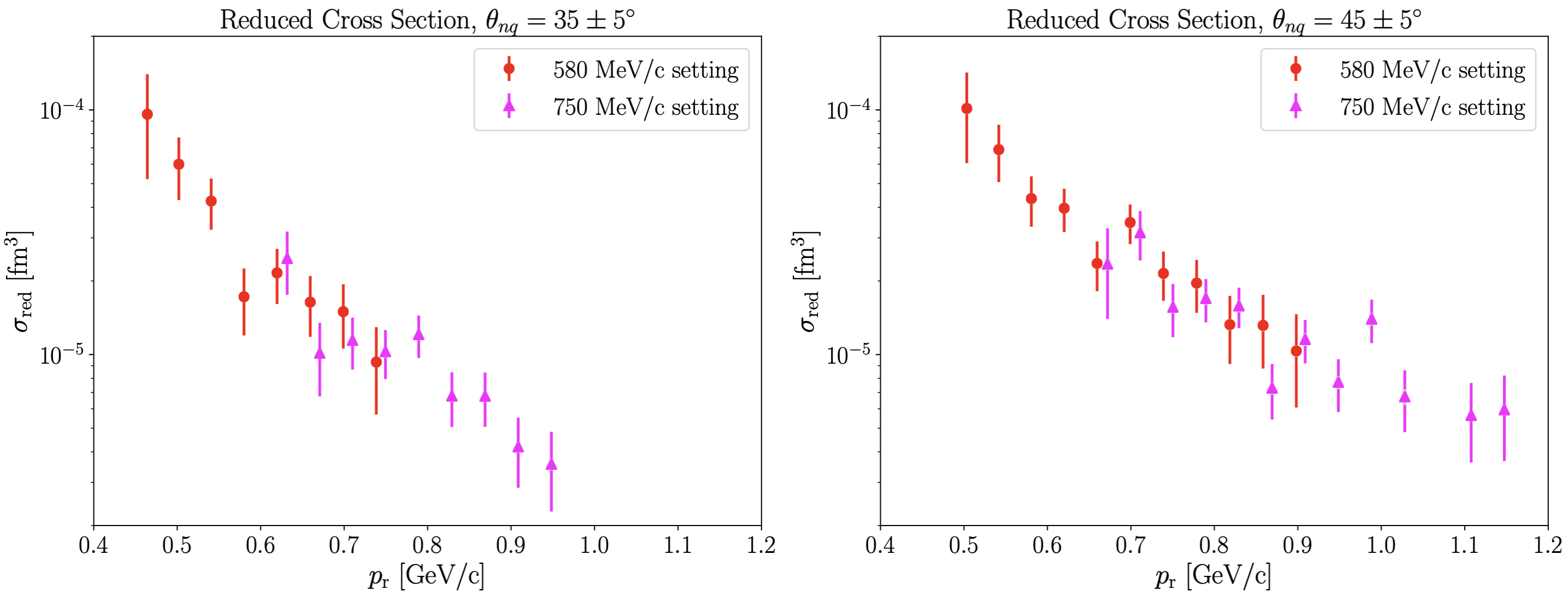}
\caption{Reduced cross section overlap region between the 580 and 750 MeV/c settings. The 750 MeV/c setting data points have been offset right by +0.01 GeV/c for ease of comparison.}
\label{fig:RedXsec_overlap}
\end{figure}

\indent The average experimental cross section was extracted by taking the ratio of the radiative corrected data yield ($Y_{\mathrm{corr}}$) to the Monte Carlo generated phase space volume for each
kinematic bin in ($\theta_{nq}$, $p_{\mathrm{r}}$). For illustration purposes, Fig. \ref{fig:Pm_Ps} shows the experimental data yield (left) and the spectrometers' phase space volume (right) binned in
missing momentum and $\theta_{nq} = 35^{\circ}$ and  $45^{\circ}$ bins for each of the three central momentum settings. Figure \ref{fig:RedXsec_overlap} shows the reduced cross section overlap for the high
$p_{\mathrm{r}}$ region. A detailed discussion of how the experimental and reduced cross sections were extracted can be
found in Sections 5.1 and 6.1 of Ref. \cite{cyero_phdthesis}.
\section{\large Radiative \& Bin-Centering Corrections}
\begin{figure}[!h]
\includegraphics[scale=0.27]{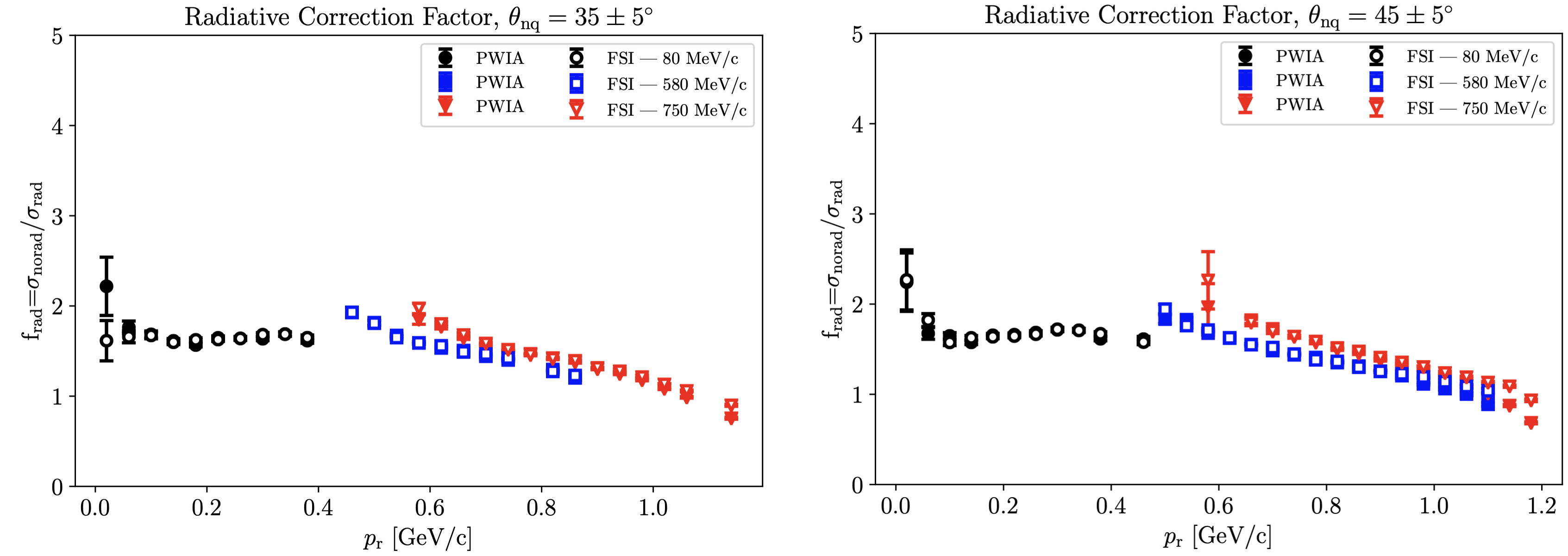}
\caption{Radiative correction factor versus neutron recoil momenta, $p_{\mathrm{r}}$, for $\theta_{nq}=35^{\circ}$ (left) and $45^{\circ}$ (right). }
\label{fig:RC_factor}
\end{figure}

\begin{figure}[!ht]
\includegraphics[scale=0.27]{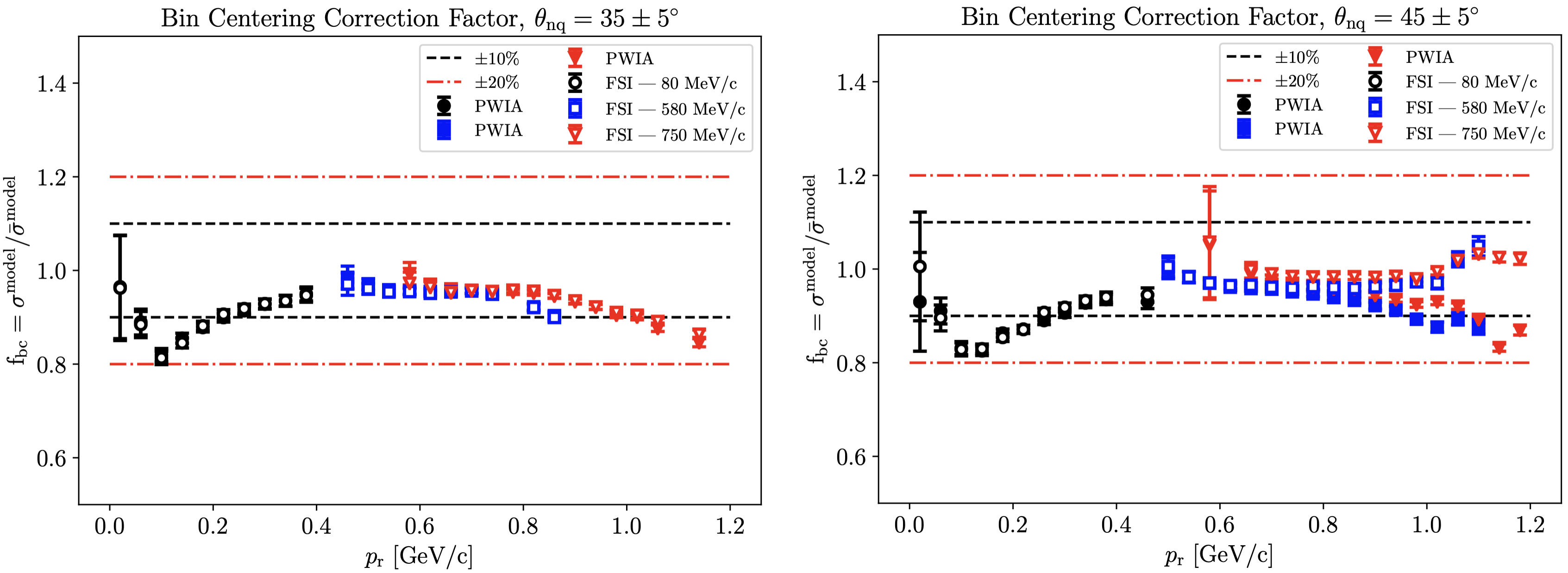}
\caption{Bin centering correction factor versus neutron recoil momenta, $p_{\mathrm{r}}$, for $\theta_{nq}=35^{\circ}$ (left) and $45^{\circ}$ (right).
  The inner (black dashed) and outer (red dash-dotted) lines represent a percent deviation from unity of $\pm10\%$ and $\pm20\%$, respectively.}
\label{fig:BC_factor}
\end{figure}

\indent The radiative corrections were applied by multiplying the ratio of non-radiative to radiative SIMC yields to the data yield for each ($\theta_{nq}$, $p_{\mathrm{r}}$)
kinematic bin as described in the Letter. The radiative correction factors for $\theta_{nq}=35^{\circ}$ and $45^{\circ}$ are shown in Fig. \ref{fig:RC_factor}. The calculation was done using the
Laget PWIA and FSI models for systematic effect studies, but the FSI model was ultimately used to correct the data yield. \\
\indent The uncertainty in the radiative
correction factor was determined for each ($\theta_{nq}$, $p_{\mathrm{r}}$) bin by the Monte-Carlo statistics from the simulation and was propagated to the final cross section error.
The large uncertainty in the lowest momentum bin, $p_{\mathrm{r}} = 0.02\pm0.02$, can be understood from the fact that statistics were limited in this phase space region as indicated by
the "hole" observed at $p_{\mathrm{r}} = 0.02$ in Fig. \ref{fig:Pm_Ps} (right). 
\\
The bin centering corrections were applied by multiplying the ratio, $f_{\mathrm{bc}} \equiv \sigma_{\mathrm{avg.kin}}/\bar{\sigma}$, to the average data cross section over each ($\theta_{nq}$, $p_{\mathrm{r}}$)
kinematic bin, as described in the Letter. The bin centering correction factors for $\theta_{nq}=35^{\circ}$ and $45^{\circ}$ are shown in
Fig. \ref{fig:BC_factor}. The calculation was done using the Laget PWIA and FSI models for systematic effect studies, but the FSI model was ultimately used to correct the data cross section. \\
\\
\indent The uncertainty in the bin-centering correction factor was determined for each  ($\theta_{nq}$, $p_{\mathrm{r}}$) bin
using the standard error propagation for a ratio of two quantities and the result was propagated to the final cross section error. \\
\section{\large Systematic Uncertainty Studies on Radiative and Bin-Centering Corrections }
\begin{figure}[!h]
\includegraphics[scale=0.37]{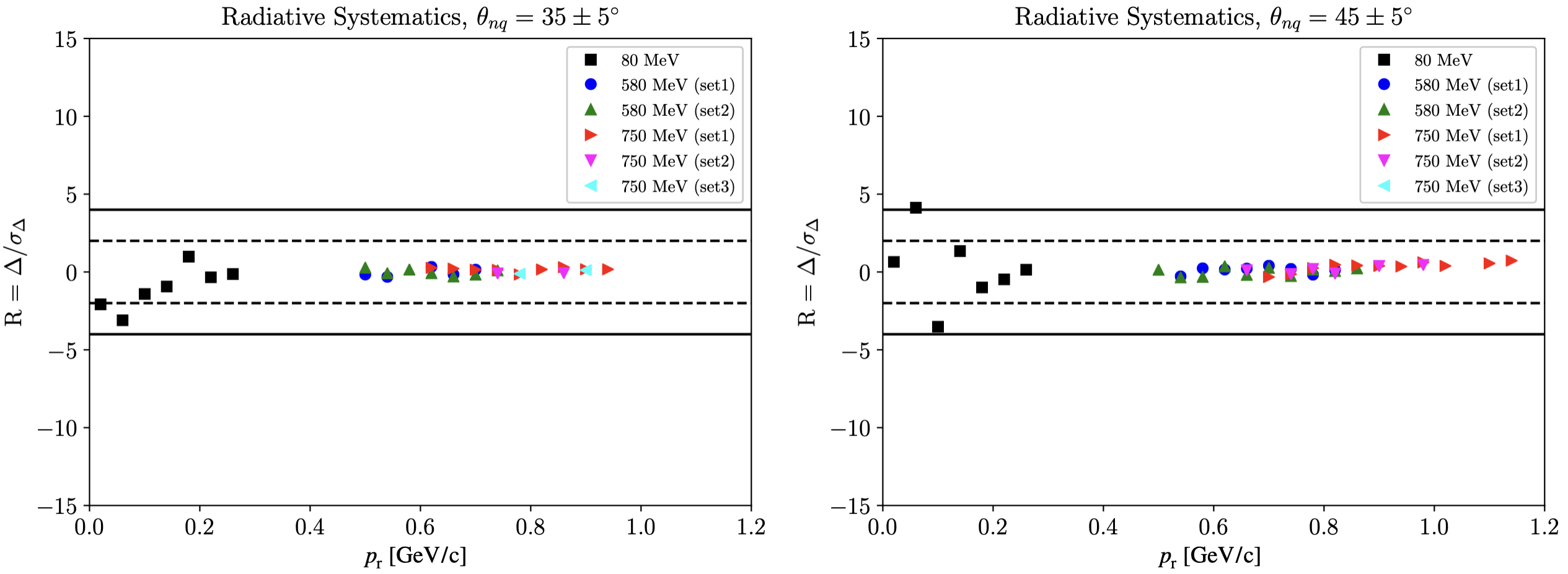}
\caption{Systematic effects of the radiative corrections model dependency on the data cross sections
  for $\theta_{nq}=35^{\circ}$ (left) and $45^{\circ}$ (right). The inner (black dashed) and outer (black solid)
  lines represent the $\Delta=\pm2\sigma_{\Delta}$ and $\pm4\sigma_{\Delta}$ boundaries, respectively.  }
\label{fig:rad_sys}
\end{figure}

\begin{figure}[!ht]
\includegraphics[scale=0.37]{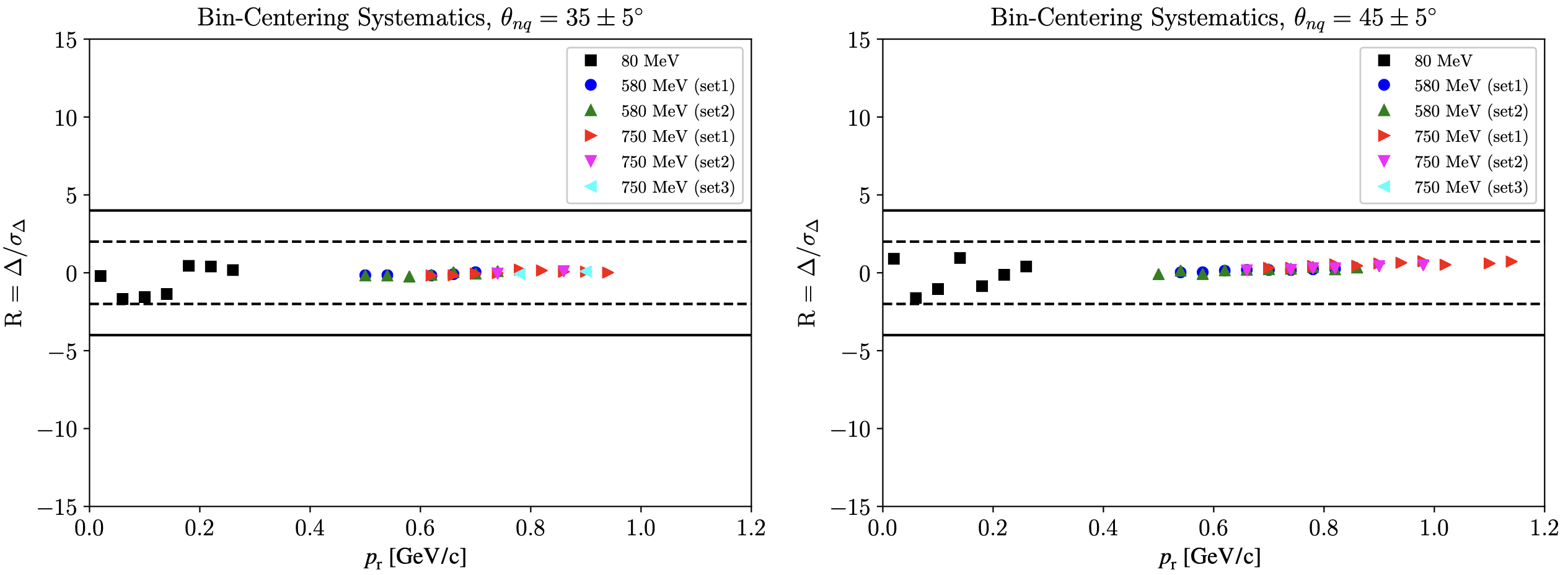}
\caption{Same as Fig. \ref{fig:rad_sys} but for bin-centering correction systematics.}
\label{fig:bc_sys}
\end{figure}

\indent The systematic effect on the cross sections due to model dependency of the radiative and bin-centering corrections was investigated (see Section 5.10 of Ref. \cite{cyero_phdthesis})
using the Barlow ratio approach \cite{barlow2002systematic,barlow2017}. In this case, the ratio was calculated from the difference between the experimental cross
sections using the Laget PWIA and FSI models for both radiative and bin-centering corrections. Figures \ref{fig:rad_sys} and \ref{fig:bc_sys}
show the Barlow ratio for radiative and bin-centering systematics is mostly within 2 standard deviations which show that
the model dependency of the correction factors have a negligible effect on the experimental cross sections.
\\
\indent Systematic studies of the event selection cuts described above were also carried out using the R. Barlow approach. The studies demonstrated that variations in the software had a negligible
effect in the measured cross sections and thus no systematic uncertainties due to software cuts were included in the final result. For a detailed discussion of these studies see Section 5.10 of Ref. \cite{cyero_phdthesis}.
\\\\
\bibliography{supp}